\documentclass[10pt,english,journal]{IEEEtran}
\usepackage[T1]{fontenc}
\usepackage[utf8]{inputenc}
\usepackage{geometry}
\usepackage{amsmath}
\usepackage{xpatch}
\usepackage{amssymb}
\usepackage{mathtools}
\usepackage{amsthm}
\usepackage{nicefrac}
\usepackage{graphicx}
\usepackage{dsfont}
\usepackage{array,ragged2e}
\usepackage{enumitem}
\usepackage{etoolbox}
\usepackage{babel}
\usepackage[ruled,vlined,linesnumbered]{algorithm2e}
\usepackage{balance}
\usepackage{float}
\usepackage{hyperref}
\usepackage[usenames,dvipsnames,table]{xcolor}
\usepackage{subcaption}
\usepackage{listings}
\usepackage{booktabs}
\usepackage{multirow}
\usepackage{tabularx}
\usepackage{cite}
\usepackage{url}
\usepackage{pifont}

\definecolor{lightgraybg}{RGB}{240,240,240}
\definecolor{revblue}{RGB}{0,0,190}

\newcommand{\rev}[1]{#1}

\usepackage[framemethod=default]{mdframed}
\mdfdefinestyle{thmbox}{%
  linewidth=0.6pt, linecolor=black!30, backgroundcolor=black!4,
  innertopmargin=3pt, innerbottommargin=3pt,
  innerleftmargin=5pt, innerrightmargin=5pt,
  skipabove=3pt, skipbelow=3pt, splittopskip=5pt}

\theoremstyle{plain}
\newtheorem{theorem}{Theorem}

\theoremstyle{definition}
\newtheorem{definition}{Definition}
\newtheorem{assumption}{Assumption}

\surroundwithmdframed[style=thmbox]{theorem}
\surroundwithmdframed[style=thmbox]{definition}

\DeclareMathOperator*{\argmax}{arg\,max}
\DeclareMathOperator*{\argmin}{arg\,min}
\DeclareMathOperator{\sd}{sp}
\newcommand{\diag}{\mathop{\mathrm{diag}}}
\newcommand{\KL}{D_{\mathrm{KL}}}
\newcommand{\Mem}{\mathcal{M}}
\newcommand{\Prob}{\mathbb{P}}
\newcommand{\Exp}{\mathbb{E}}

\graphicspath{ {Figures/} }

\lstdefinestyle{promptstyle}{
    backgroundcolor=\color{gray!10},
    basicstyle=\ttfamily\scriptsize,
    breaklines=true, columns=fullflexible,
    showstringspaces=false, numbers=none, frame=none,
    captionpos=b, tabsize=2
}
\begin{document}

\title{Toward an Unbiased Collective Memory for Efficient LLM-Based Agentic 6G Cross-Domain Management}

\author{
Hatim~Chergui,~\IEEEmembership{Senior~Member,~IEEE},
Farhad~Rezazadeh,~\IEEEmembership{Member,~IEEE},
Miguel Catalan Cid, \\ Pouria Sayyad Khodashenas, Daniel Camps Mur, and Christos~Verikoukis,~\IEEEmembership{Senior~Member,~IEEE}
\IEEEcompsocitemizethanks{\IEEEcompsocthanksitem Hatim Chergui, Miguel Catalan Cid, Pouria Sayyad Khodashenas and Daniel Camps Mur are with the i2CAT Foundation, Spain. (e-mails: chergui@ieee.org, name.surname@i2cat.net)}
\IEEEcompsocitemizethanks{\IEEEcompsocthanksitem F. Rezazadeh is with the BrainOmega and the Technical University of Catalonia (UPC), 08028 Barcelona, Spain (e-mail: farhad.rezazadeh@upc.edu).}
\IEEEcompsocitemizethanks{\IEEEcompsocthanksitem Christos Verikoukis is with ISI/ATH and University of Patras, Greece. (e-mail: cveri@isi.gr)}}
\maketitle

\begin{abstract}
Agentic artificial intelligence is a candidate enabler of Level-4 autonomy in sixth-generation (6G) networks, but agents reasoning over a shared memory inherit its distortions. We study cross-domain radio access network (RAN)--edge orchestration in which a RAN agent minimizing energy and an edge agent minimizing latency negotiate, validate proposals against a digital twin (DT), and share a collective memory. Such a memory is a sampling device: what agents retrieve, not what they store, determines what they do. We define the Retrieval Bias Index, a divergence between the induced retrieval law and a representative reference, decomposed along temporal, confirmation and availability axes, and show that additive retrieval scoring is an exponential tilt of the undebiased law, yielding a closed form for the failure-amplification weight and a saturation floor set by the store's own composition; a further result bounds excess service level agreement (SLA) violation additively in retrieval bias and DT mismatch. On a tandem queue for which stochastic network calculus supplies a feasible region, an oracle and a Nash bargaining reference, the operator attains a $12.7\times$ bias reduction over a recency store and the smallest distance to the bargaining solution among six memory architectures.
\end{abstract}


\section{Introduction}
\IEEEPARstart{T}{raditional} telecom automation---spanning TM Forum Levels~1 to~3---has delivered incremental efficiency through scripting, analytics and closed-loop control, but remains task-specific, reactive, and fragmented across vendor-specific platforms and loosely integrated toolchains. This fragmentation constrains the ability to generalize, collaborate or adapt in real time, blocking the leap toward higher autonomy. Reaching Level~4 requires intelligence that is connected, semantic and proactive: agents that perceive through (multimodal) telemetry, reason over past experience, plan, communicate under explicit intents, and act by invoking tools and application programming interfaces (APIs). Such agents can negotiate across the radio access network (RAN), edge and core silos and reconcile objectives that are structurally in conflict---energy on one side, latency on the other.

\rev{Memory is what separates an agent from a stateless controller. An agent that logs every negotiation, distills each into a context--action--outcome record and consults those records before proposing a configuration is performing a form of non-parametric policy improvement, in which the store stands in for an explicit model of the environment. But this only works if the records it retrieves are representative of the records it holds. Retrieval is a sampling operator, and every practical retrieval operator is skewed: recency weighting over-samples the last few episodes, similarity search over-samples one cluster of the context space, and---most damaging in a control loop---success-conditioned relevance under-samples the failures from which the agent has the most to learn. These are the machine analogues of recency, availability and confirmation bias~\cite{kahneman,gilovichs}, and in a network control loop they are not a curiosity: they are the mechanism by which a nominally converged agent drifts into SLA violation, because the evidence on which it conditions has quietly stopped resembling the conditions it faces. The drift is moreover self-reinforcing, since the store is written by the same policy that reads it: a retrieval rule that under-samples failures yields configurations whose failures the store then under-records, so the skew deepens over a deployment rather than averaging out. In the slice considered here each distortion has a concrete operational form. Recency weighting recalls a run of quiet low-traffic episodes and forgets the bandwidth reservation that failed during the last arrival burst; similarity search returns the cluster of medium-traffic precedents that dominate the store while the rare deep-fade episodes go unretrieved; and success-conditioned relevance surfaces the configurations that met the SLA while omitting the leaner ones that breached it, which are the only records locating the boundary. The pathology is moreover a property of the retrieval operator rather than of the store, since a representative store queried by a skewed operator still supplies skewed evidence.}

\rev{Consider a network slice spanning a RAN and an edge site, in which a RAN agent lowers bandwidth to save energy while an edge agent raises CPU frequency to protect latency, subject to a tail constraint $\Prob(L>L_{\max})\le\epsilon$ characteristic of ultra-reliable low-latency communication (URLLC). The joint configuration is not separable, since the delay budget is shared across the computation and transmission stages, so any reduction in bandwidth must be compensated in CPU frequency and vice versa; neither agent observes the other's utility, and both act through a language model reasoning over a shared memory. The setting is therefore adversarial in objective but cooperative in constraint, and the agents' only common ground is the record of what has previously worked --- which is exactly the object whose skew we are concerned with. Three questions follow, none of which the literature answers: i)~what is the \emph{correct} operating point, so that agentic performance can be measured as a gap rather than asserted as an improvement? ii)~what, quantitatively, is a biased memory, and can that quantity be related to a network KPI? iii)~can the mitigation weights be derived rather than tuned?}

\subsection{Contributions}
\rev{Against the state of the art surveyed in Section~\ref{sec:related}, our contributions are:}
\begin{itemize}[leftmargin=*,itemsep=1pt,topsep=2pt]
\item \rev{a tandem RAN--edge model with Markov-modulated Poisson arrivals, a finite-state Markov fading channel and HARQ, for which Theorem~\ref{thm:snc} yields a closed-form delay-violation bound and, from it, the SLA-feasible region, an oracle allocation and a unique Nash bargaining solution, so agentic performance is measured as a gap from a known optimum;}
\item \rev{the Retrieval Bias Index (Definition~\ref{def:rbi}), a divergence between the induced retrieval law and a representative reference, decomposed into temporal, confirmation and availability components, computable from the agents' own logs and independent of the reasoner that consumes it;}
\item \rev{Theorem~\ref{thm:tilt}, which shows that additive retrieval scoring is an exponential tilt and so replaces hyperparameter search with the closed form $\delta^\star=\log(\rho_S/\rho_F)$, together with a saturation floor showing that confirmation bias is correctable only insofar as failures are retained;}
\item \rev{Theorem~\ref{thm:master}, which bounds excess SLA violation by $\sqrt{\mathcal{B}/2}+\sqrt{\varepsilon_{\mathrm{DT}}/2}$, separating memory bias from digital-twin mismatch additively, so that observed degradation can be attributed to a stale memory or a drifting twin;}
\item \rev{a semantic--relational memory realizing the operator, combining hybrid dense, reranked, sparse and numeric retrieval with grounded admission, Beta consolidation and stratified eviction, and evaluated against a dense-retrieval baseline sharing the identical stack without the tilt, which isolates debiasing from better semantics.}
\end{itemize}

The remainder is organized as follows. Section~\ref{sec:related} surveys related work and positions our contribution. Section~\ref{sec:model} develops the system model and the optimization problem, and establishes the delay-violation region and the bargaining reference. Section~\ref{sec:framework} details the agentic negotiation framework. Section~\ref{sec:memory} defines retrieval bias and presents the unbiased collective memory with its guarantees. Section~\ref{sec:results} reports the evaluation and Section~\ref{sec:conclusion} concludes.

\section{Related Work and Positioning}
\label{sec:related}
\begin{table*}[!t]
\centering
\rev{
\caption{Positioning with respect to representative prior art.}
\label{tab:related}
\small
\setlength{\tabcolsep}{5pt}
\renewcommand{\arraystretch}{1.1}

\begin{tabular}{lccccccc}
\toprule
\textbf{Work}
& \textbf{Domain}
& \textbf{Agent}
& \textbf{Structured}
& \textbf{Bias}
& \textbf{Derived}
& \textbf{KPI}
& \textbf{Open} \\
&
& \textbf{memory}
& \textbf{memory}
& \textbf{metric}
& \textbf{weights}
& \textbf{bound}
& \textbf{model} \\
\midrule

LINKs~\cite{jiang2024links}
& 6G/DT
& \ding{55}
& \ding{55}
& \ding{55}
& \ding{55}
& \ding{55}
& \ding{55} \\

O-RAN agents~\cite{tang2025edge}
& O-RAN
& \ding{55}
& \ding{55}
& \ding{55}
& \ding{55}
& \ding{55}
& \ding{55} \\

MemGPT~\cite{packer2023memgpt}
& General
& \ding{51}
& Tiered
& \ding{55}
& \ding{55}
& \ding{55}
& \ding{51} \\

Mem0~\cite{chhikara2025mem0}
& General
& \ding{51}
& Graph
& \ding{55}
& \ding{55}
& \ding{55}
& \ding{51} \\

A-MEM~\cite{xu2025amem}
& General
& \ding{51}
& Graph
& \ding{55}
& \ding{55}
& \ding{55}
& \ding{51} \\

Gen. Agents~\cite{park2023generative}
& Simulation
& \ding{51}
& Flat
& \ding{55}
& \ding{55}
& \ding{55}
& \ding{55} \\

Reflexion~\cite{shinn2023reflexion}
& General
& \ding{51}
& Episodic
& \ding{55}
& \ding{55}
& \ding{55}
& \ding{51} \\

\textcolor{black}{MemoryBank~\cite{zhong2024memorybank}}
& \textcolor{black}{Dialogue}
& \textcolor{black}{\ding{51}}
& \textcolor{black}{Hierarchical}
& \textcolor{black}{\ding{55}}
& \textcolor{black}{\ding{55}}
& \textcolor{black}{\ding{55}}
& \textcolor{black}{\ding{51}} \\

\textcolor{black}{Memory-R1~\cite{yan2025memoryr1}}
& \textcolor{black}{Dialogue}
& \textcolor{black}{\ding{51}}
& \textcolor{black}{Flat bank}
& \textcolor{black}{\ding{55}}
& \textcolor{black}{\ding{55}}
& \textcolor{black}{\ding{55}}
& \textcolor{black}{\ding{51}} \\

MindScope~\cite{mindscope}
& General
& \ding{55}
& \ding{55}
& Dataset
& \ding{55}
& \ding{55}
& \ding{51} \\

Anchoring~\cite{chergui2026anchoring}
& 6G
& \ding{51}
& Flat
& Partial
& \ding{55}
& \ding{55}
& \ding{55} \\

\textbf{This work}
& 6G RAN--Edge
& \ding{51}
& Graph
& \textbf{RBI}
& \ding{51}
& \ding{51}
& \ding{51} \\

\bottomrule
\end{tabular}
}
\end{table*}

\subsection{Agentic AI for Autonomous Networks}
LINKs~\cite{jiang2024links} couples LLMs with digital-twin-based 6G management, enabling agents to interpret intent and issue control actions, though its reasoning depth and coupling to real-time RAN telemetry remain limited. \cite{xia2025agentic} conceptualizes generative models as distributed agents performing collaborative planning across network elements, and LaMeta~\cite{liu2025lameta} decouples cloud-side intent parsing from edge-side policy generation; both are validated only in high-level simulation. \cite{xu2025urban} explores agentic digital twins for cyber-physical optimization in urban logistics, \cite{zhang2025toward} integrates structured network knowledge into the reasoning loop to curb hallucination, and \cite{tang2025edge} outlines an architecture for embedding LLM agents in O-RAN xApp and rApp lifecycle management. \rev{More recently, \cite{rezazadeh2025worldmodel} moves agentic inference toward the near-RT RIC timescale. Common to all of these is that memory, where present, is an undifferentiated retrieval backend; none asks whether what is retrieved is representative of what is stored.}



\subsection{Memory Systems for LLM Agents}
\label{sec:relmem}

\rev{Outside telecom, agent memory is an active design space.
MemGPT~\cite{packer2023memgpt} borrows the virtual-memory metaphor,
paging between main, recall and archival context under
interrupt-driven control, with no principled forgetting.
Mem0~\cite{chhikara2025mem0} implements an
extraction--consolidation--retrieval pipeline but delegates
retention decisions to the LLM.
A-MEM~\cite{xu2025amem} organizes memories as a self-linking
Zettelkasten network, generating structured attributes per note
and letting the model decide how notes link and evolve, which
yields substantially better recall across six foundation models.
Generative Agents~\cite{park2023generative} scores memories by a
weighted sum of recency, importance and relevance, the de facto
default and our strongest baseline, and
Reflexion~\cite{shinn2023reflexion} converts episode outcomes into
verbal self-critiques.
MemoryBank~\cite{zhong2024memorybank} introduces
explicit forgetting through an Ebbinghaus-inspired exponential
decay mechanism, strengthening memories when they are recalled
to support persistent, personalized interaction.
Memory-R1~\cite{yan2025memoryr1} learns memory management through
reinforcement learning; a Memory Manager selects operations to
add, update, delete or retain entries, while an Answer Agent
filters retrieved memories and reasons over them, with both
agents trained using rewards tied to answer correctness.}

\rev{These systems emphasize useful recall,
coherent interaction, or downstream task performance.
Such objectives do not, by themselves, ensure the
\emph{statistical representativeness} of the retrieved set
with respect to the operating distribution.
This distinction matters in a control loop, where retrieved
experiences influence the policy and an unrepresentative sample
can distort the evidence used for decision-making. In particular, Generative-Agents scoring favors recently accessed memories through recency weighting and semantically
similar memories through relevance weighting. These mechanisms can alter the distribution of retrieved
experiences; a further success bias can arise when memory admission or retrieval preferentially selects successful outcomes. Thus, retrieval utility and statistical representativeness are distinct design objectives, and improvements in the former do not establish the latter.}

\subsection{Cognitive Distortions in LLM Multi-Agent Systems}
MindScope~\cite{mindscope} contributes a dataset of 72 human cognitive biases and shows that latent distortions such as anchoring surface in agent--agent interaction even when individual models are aligned. \cite{unmasking} shows that iterative multi-agent discussion forms echo chambers that amplify bias through consensus seeking, \cite{fairness} links ethical alignment to incentive design, and the Hidden Profile benchmark~\cite{assessing} shows agent teams failing to surface information held by a single peer. \rev{More recent work reports that multi-agent structure can \emph{amplify} rather than mitigate bias~\cite{emergentbias2025,oh2025debate}. Within telecom, \cite{chergui2025tutorial} catalogues cognitive biases in agentic 6G networks and \cite{chergui2026anchoring} targets anchoring bias specifically.}

\rev{The prevailing mitigation strategy there is \emph{dialogical} --- debate, self-consistency, persona diversity --- whereas our position is that in a memory-equipped system the dialogical layer is downstream of the sampling layer, and that debiasing the sampling operator is both cheaper and provably characterizable, while the effect of debate is empirical and, per \cite{emergentbias2025,oh2025debate}, not reliably signed. Table~\ref{tab:related} summarizes the comparison: to the best of our knowledge this is the first work to supply a computable metric of retrieval bias for an agent memory, to derive rather than tune the corresponding mitigation weights, to bound a network KPI in terms of that metric, and to do so inside a cross-domain 6G control loop with an analytically characterized optimum.}

\begin{figure}[!t]
    \centering
    \includegraphics[width=\linewidth]{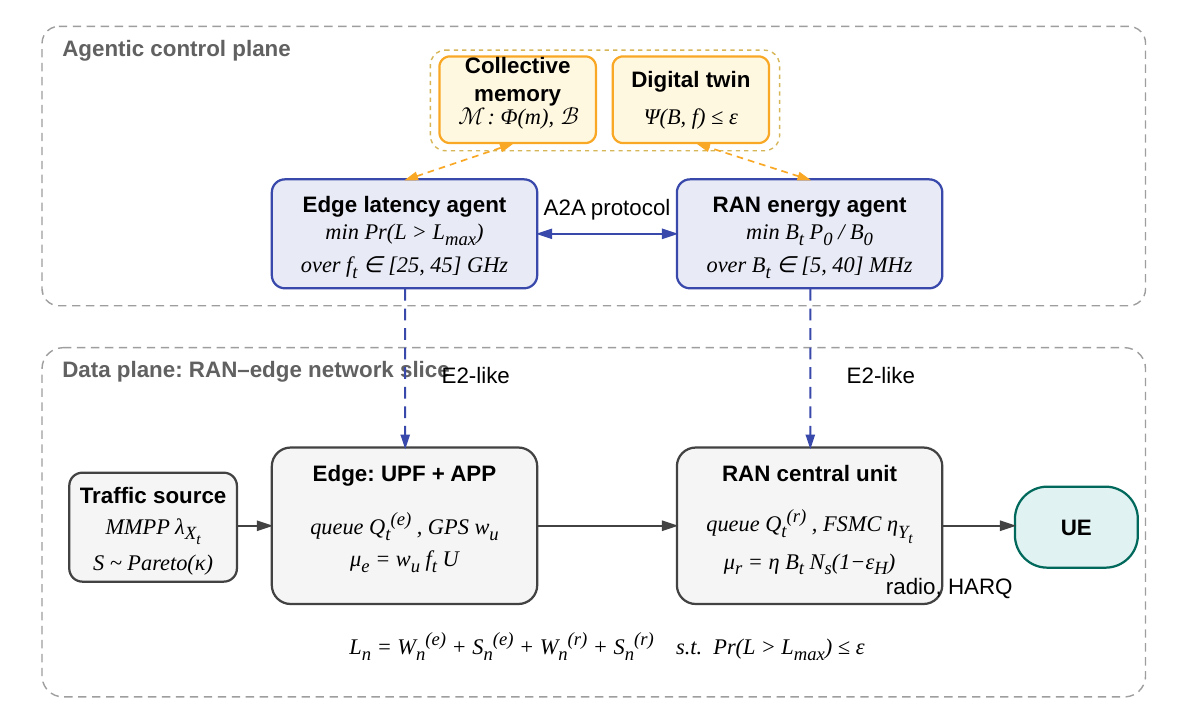}
    \caption{Cross-domain RAN--edge system model. The agentic control plane
    (top) holds the collective memory and the digital twin as shared tools; the
    edge latency agent and the RAN energy agent negotiate over the A2A protocol
    and enforce the agreed configuration through an E2-like interface.}
    \label{fig:arch}
\end{figure}

\section{System Model and Problem Formulation}
\label{sec:model}

\subsection{Network and Traffic Model}
Figure~\ref{fig:arch} depicts a 6G network slice spanning the RAN and edge domains. The edge domain hosts a containerized application server (APP) and a user-plane function (UPF), both deployed as cloud-native functions (CNFs) on a platform of maximum aggregate frequency $f_{\max}$, processing $U$ bits per CPU cycle per Hz. The RAN central unit (CU) transmits over the radio interface. Each domain is managed by an autonomous LLM-based agent with a distinct objective: the RAN agent minimizes energy by adjusting the allocated bandwidth $B_t$, the edge agent minimizes end-to-end delay by tuning the CPU frequency $f_t$ allocated to the UPF. Agreements are enforced over an E2-like interface into the live environment.

\rev{A simpler formulation --- deterministic fluid queues with independent and identically distributed (i.i.d.) uniform spectral efficiency, and latency taken from the time-averaged form of Little's law --- is not adequate here. Such a model has no delay tail, which makes a tail SLA unmeasurable, precludes any study of long-tail anomalies, and leaves the digital twin nearly identical to the environment so that model mismatch cannot be exercised. We therefore adopt the following stochastic model, chosen so that it is simultaneously (i) rich enough to exhibit burstiness, correlated fading and retransmission dynamics, and (ii) analytically tractable through effective-bandwidth theory, which is what supplies the optimum against which the agents are measured.}

\rev{Turning to the load, downlink requests arrive at the APP according to a Markov-modulated Poisson process (MMPP). Let $\{X_t\}$ be an irreducible aperiodic Markov chain on $\mathcal{X}=\{\text{low},\text{medium},\text{high}\}$ with transition matrix $\mathbf{P}_A$ and stationary law $\boldsymbol{\pi}_A$. Conditioned on $X_t=i$, the number of packets arriving in slot $t$ of duration $\tau$ is Poisson with mean $\lambda_i\tau$, and each packet carries a payload $S$ drawn from a heavy-tailed law}
\begin{equation}
\Prob(S>s)=\left(\frac{s_0}{s}\right)^{\kappa}\!\!,\quad s_0\le s\le s_{\max},\ \ \kappa\in(1,2],
\label{eq:pareto}
\end{equation}
\rev{so that the arrival process exhibits both short-range Markov burstiness and heavy-tailed batch sizes. The truncation at $s_{\max}=40\,\Exp[S]$ is not cosmetic: an untruncated Pareto law with $\kappa\le 2$ has no finite moment generating function, so the effective-bandwidth machinery of Section~\ref{sec:region} does not apply to it and the bound of Theorem~\ref{thm:snc} is violated outright, as we confirmed in simulation before imposing the truncation. A bounded payload is also the physically correct model, since a burst cannot exceed the buffer it is drawn from. The bit arrival volume in slot $t$ is $U_t^{(a)}=\sum_{n=1}^{N_t}S_n$ with $N_t\sim\mathrm{Poisson}(\lambda_{X_t}\tau)$. The modulating state $X_t$ is the latent variable that the agents' memory records as \texttt{traffic\_level}; it is now a genuine hidden state rather than a post-hoc label, which is what makes context-conditioned retrieval meaningful.}

\subsection{Service Stages and Power Consumption}
\rev{The UPF and APP share the edge platform under weighted fair queueing (WFQ) with weights $(w_u,w_a)$, $w_u+w_a=1$: each backlogged class receives compute in proportion to its weight, so the UPF is served at an instantaneous rate $w_u f_t U$ bits/s. This is what makes the edge a \emph{multi-class} server rather than a dedicated one --- the UPF cannot claim the whole platform while the application server is busy --- and it is the discipline a container runtime implements through CPU shares. Per-packet service times are lognormal, $\ln S^{\mathrm{svc}}\sim\mathcal{N}(m_c,s_c^2)$, capturing container scheduling jitter, interrupt coalescing and cache effects, so that the stage is a multi-class $M/G/1$ queue rather than a deterministic server. The computation backlog evolves as}
\begin{equation}
Q_{t+1}^{(e)}=\big(Q_t^{(e)}-U_t^{(e)}\big)^{+}+U_t^{(a)},\qquad U_t^{(e)}=\tau\,w_u f_t U\,\xi_t^{(e)},
\end{equation}
\rev{where $(\cdot)^{+}=\max(\cdot,0)$ and $\xi_t^{(e)}$ is the unit-mean lognormal service-rate perturbation.}

\rev{At the second stage, spectral efficiency is generated by a finite-state Markov channel (FSMC) rather than sampled i.i.d.\ per slot: the fading envelope is partitioned into $M$ intervals, each becoming a state with a representative spectral efficiency, and transitions are permitted between adjacent states only. Let $\{Y_t\}$ be a Markov chain on $\mathcal{Y}=\{\eta_1,\dots,\eta_{M}\}$, obtained by partitioning the Rayleigh/Rician envelope into $M$ regions with transition matrix $\mathbf{P}_C$ built from the level-crossing rates at Doppler frequency $f_D$. This introduces temporal correlation in the channel, and hence deep-fade episodes of realistic duration---the regime in which the agents' proposals actually fail. The nominal per-slot transmission volume is}
\begin{equation}
U_t^{(r)}=\tau\,\eta_{Y_t}B_tN_s\big(1-\varepsilon_{\mathrm{H}}(Y_t)\big),
\end{equation}
\rev{where $N_s\sim\mathcal{O}(\min(N_t,N_r))$ is the spatial multiplexing gain and $\varepsilon_{\mathrm{H}}(j)$ is the residual error probability in fading state $j$ after $R_{\max}$ rounds, so that retransmissions inflate the effective service requirement. Taking retransmissions to fail independently, $\varepsilon_{\mathrm{H}}(j)=p_1(j)^{R_{\max}}$ with a first-transmission error probability $p_1(j)=p_0\big(1-\tfrac{j}{M-1}\big)$ that decays linearly across the FSMC states, $p_0$ being the packet error rate in the deepest fade. This is a deliberately conservative model: it neglects the soft-combining gain of incremental redundancy, so the transmission stage it induces is no faster than a real one. The CU backlog is}
\begin{equation}
Q_{t+1}^{(r)}=\big(Q_t^{(r)}-U_t^{(r)}\big)^{+}+\min\big(Q_t^{(e)},U_t^{(e)}\big).
\end{equation}

Finally, with reference power $P_0$ per $B_0=20$~MHz carrier, the RAN power at slot $t$ is $P_t=B_tP_0/B_0$, and the realized energy saving relative to the maximum-bandwidth configuration is $1-B_t/B_{\max}$.

\subsection{End-to-End Delay and Problem Formulation}
\rev{Rather than converting a time-averaged backlog into a mean delay, we track per-packet sojourn times directly by the Lindley recursion at each stage,}
\begin{equation}
W_{n+1}^{(k)}=\big(W_n^{(k)}+S_n^{(k)}-A_n^{(k)}\big)^{+},\qquad k\in\{e,r\},
\label{eq:lindley}
\end{equation}
\rev{with $S_n^{(k)}$ the service time of packet $n$ at stage $k$ and $A_n^{(k)}$ its interarrival time, so that the end-to-end delay of packet $n$ is}
\begin{equation}
L_n=W_n^{(e)}+S_n^{(e)}+W_n^{(r)}+S_n^{(r)}.
\label{eq:e2e}
\end{equation}
\rev{The quantity of interest is then the \emph{violation probability} $\Prob(L>L_{\max})$, not $\Exp[L]$. This distinction matters: in our traces the mean delay of a configuration and its violation probability are only weakly related, and it is the latter that the SLA constrains.}

\rev{With the delay defined, the cross-domain resource allocation problem is the chance-constrained program}
\begin{equation}
\begin{aligned}
(\mathcal{P}1):\quad \min_{B\in\mathcal{B},\,f\in\mathcal{F}}\ \ & \frac{B\,P_0}{B_0}\\
\text{s.t.}\ \ & \Prob\big(L>L_{\max}\big)\le\epsilon,\\
& B\le B_{\max},\quad f\le f_{\max}.
\end{aligned}
\label{eq:P1}
\end{equation}
\rev{$(\mathcal{P}1)$ is not solvable in closed form by either agent alone: the RAN agent controls $B$ but is constrained by a quantity that depends on $f$, and symmetrically for the edge agent, while neither observes the other's utility. This is precisely the structure that motivates negotiation rather than centralized optimization. What we can do, however, is characterize the constraint set analytically, which turns the negotiation into a problem with a known answer.}

\subsection{Delay-Violation Region}
\label{sec:region}
\rev{We invoke effective-bandwidth and effective-capacity theory. For the modulated arrival process, define}
\begin{equation}
\begin{split}
\alpha_A(\theta)&=\tfrac{1}{\theta}\log\sd\!\Big(\mathbf{P}_A\,\diag\big(e^{\Lambda_i(\theta)}\big)\Big),\\
\Lambda_i(\theta)&=\lambda_i\tau\big(M_S(\theta)-1\big),
\end{split}
\label{eq:effbw}
\end{equation}
\rev{where $M_S(\theta)=\Exp[e^{\theta S}]$ is the payload moment generating function, finite by the truncation in \eqref{eq:pareto}, and $\sd(\cdot)$ is the spectral radius. $\Lambda_i$ is the per-slot cumulant generating function of the compound Poisson batch in modulating state $i$. Substituting the per-state \emph{mean} rate $r_i=\lambda_i\Exp[S]$, as a fluid model would, discards the batch variability and produces a bound that the simulation violates; the compound form is what makes Theorem~\ref{thm:snc} hold. Symmetrically, the effective capacity of stage $k$ is}
\begin{equation}
\alpha^{(k)}_C(\theta)=-\tfrac{1}{\theta}\log\sd\!\big(\mathbf{P}_k\,\diag(e^{-\theta c^{(k)}_j})\big),
\label{eq:effcap}
\end{equation}
\rev{with $c^{(e)}_j$ the state-dependent computation rate and $c^{(r)}_j=\eta_jBN_s(1-\varepsilon_{\mathrm{H}}(j))$ the state-dependent transmission rate. Both spectral radii are simple by Perron--Frobenius, so $\alpha_A$ and $\alpha_C^{(k)}$ are computable to machine precision.}

\begin{theorem}[Delay-violation region and its normative points]
\label{thm:snc}
\rev{Let $\theta_e(f)$ and $\theta_r(B)$ be the unique positive roots of $\alpha_A(\theta)=\alpha^{(e)}_C(\theta)$ and $\alpha_A(\theta)=\alpha^{(r)}_C(\theta)$, which exist and are unique whenever the corresponding stage is stable. Then:}

\rev{\emph{(i) Bound.} For every delay-budget split $u\in[0,1]$ and every horizon exceeding the aggregate service time, below which the tail is unity by construction,}
\begin{equation}
\begin{split}
\Prob\big(L>L_{\max}\big)&\le\Psi(B,f)\\
&\triangleq\inf_{u\in[0,1]}\Big[K_e e^{-\theta_e(f)\,uL_{\max}}\\
&\qquad\qquad+K_r e^{-\theta_r(B)(1-u)L_{\max}}\Big],
\end{split}
\label{eq:thm1}
\end{equation}
\rev{with $K_e,K_r$ the stage prefactors set by the stationary moment generating functions.}

\rev{\emph{(ii) Region.} $\theta_e$ is non-decreasing in $f$ and $\theta_r$ in $B$, so $\Psi$ is non-increasing in each argument and the SLA-feasible set $\mathcal{R}_\epsilon=\{(B,f):\Psi(B,f)\le\epsilon\}$ is upward-closed with a strictly decreasing boundary.}

\rev{\emph{(iii) Bargaining point.} With utilities $u_{\mathrm{R}}=1-B/B_{\max}$ and $u_{\mathrm{E}}=1-\Psi/\epsilon$ and the zero-surplus disagreement point $d=(0,0)$, the utility set over $\mathcal{R}_\epsilon$ closed under time sharing is convex and compact, and admits a unique Nash bargaining solution $(B^{\mathrm{N}},f^{\mathrm{N}})=\argmax_{\mathcal{R}_\epsilon}u_{\mathrm{R}}u_{\mathrm{E}}$.}
\end{theorem}

\begin{proof}\rev{Appendix~\ref{app:proof1}.}\end{proof}

\rev{Three consequences deserve emphasis. The infimum over $u$ is what couples the domains: the cruder exponent $\min\{\theta_e,\theta_r\}$ would decouple the constraint into $B\ge B_c$ and $f\ge f_c$ separately and make the negotiation trivial, so it is the delay \emph{budget split} that the agents implicitly bargain over. Second, $\Psi$ costs two eigenvalue evaluations and a line search, so the digital twin validates a proposal analytically rather than by Monte-Carlo rollout and leaves the negotiation's latency budget untouched. Third, the choice of $d$ is not innocuous: taking the enforced fallback $(B_{\max},f_{\max})$ as the disagreement point degenerates the problem, since that point already maximizes the edge agent's utility and leaves it no surplus anywhere in $\mathcal{R}_\epsilon$, collapsing the bargaining solution onto the oracle.}

\rev{The oracle allocation is $(B^\star,f^\star)=\argmin_{(B,f)\in\mathcal{R}_\epsilon}B$, obtained by a line search along $\partial\mathcal{R}_\epsilon$ and providing the regret reference $\mathrm{Reg}=\Exp[B_t]-B^\star$. It is, however, a centralized benchmark and therefore unfair to a negotiation between self-interested agents; the bargaining point of Theorem~\ref{thm:snc}(iii) is the appropriate normative target.}

\rev{Theorem~\ref{thm:snc}(iii) gives the evaluation its backbone: instead of reporting that memory-equipped agents reach lower latency than memory-less ones, we report the \emph{distance to the NBS}, $\|(B_t,f_t)-(B^{\mathrm{N}},f^{\mathrm{N}})\|$, which is zero for an ideal negotiator and is comparable across scenarios, traffic regimes and LLM backends.}

\section{Agentic Cross-Domain Negotiation Framework}
\label{sec:framework}

\subsection{Agent Roles and the A2A Protocol}
\label{sec:roles}
Each agent operates within a structured loop of three stages. \emph{(i) Memory query.} On receiving the counterpart's message together with current metrics, the agent queries the shared collective memory of Section~\ref{sec:memory}. \emph{(ii) Digital-twin validation.} The agent tests any candidate configuration against an internal DT; if the prediction violates the SLA or triggers a CPU over-allocation, it adjusts and re-tests until a compliant proposal is found or it concludes that none exists. \emph{(iii) Negotiation move.} The agent emits a message under the A2A protocol, either proposing a configuration, accepting the counterpart's, or declaring impasse.

That exchange is governed by a strict protocol. It is turn-based, and each message declares an intent in $\{$\texttt{PROPOSE\_ACTION}, \texttt{ACCEPT\_AGREEMENT}, \texttt{NO\_AGREEMENT\_POSSIBLE}$\}$. The first two carry a JSON payload specifying \texttt{ran\_bandwidth\_mhz}, \texttt{edge\_cpu\_frequency\_ghz} and a plain-text \texttt{reason}. A dedicated parser validates structure and flags malformed messages, which is essential to prevent silent impasse. \rev{An abridged transcript is given in Appendix~\ref{app:transcript}.}

\rev{Consensus is declared when both agents propose an identical configuration or one accepts the other's. Failure takes three distinguishable forms, which we log separately because they have different causes and different fixes: an explicit \texttt{NO\_AGREEMENT\_POSSIBLE}; exhaustion of the round budget without convergence; and a parsing failure. The last is a property of the language model rather than of the negotiation, and it becomes a first-class metric in Section~\ref{sec:results} once the backend is a 1B model.}

\rev{The twin itself is analytic: it evaluates $\Psi(B,f)$ from \eqref{eq:thm1} and acts as a validation gate, a proposal being admissible iff $\hat{\Psi}(B,f)\le\epsilon$ under its own parameters. Crucially it is \emph{not} the environment: we parameterize the mismatch by perturbing its view of the modulating chain, the fading transition matrix and the processing efficiency, and define}
\begin{equation}
\varepsilon_{\mathrm{DT}}=\KL\big(\Prob_{\mathrm{env}}(\cdot\mid B,f)\,\big\|\,\Prob_{\mathrm{DT}}(\cdot\mid B,f)\big),
\label{eq:dtmismatch}
\end{equation}
\rev{the one-step predictive divergence between environment and twin. Setting $\varepsilon_{\mathrm{DT}}=0$ recovers a perfect twin; sweeping it upward is the robustness study of Section~\ref{sec:results}-D, and Theorem~\ref{thm:master} predicts the resulting degradation.}

\subsection{Edge-Native Reasoning and Deployment}
\label{sec:deploy}
\rev{Relying on a proprietary API would compromise both reproducibility and deployability. The framework is instead instantiated with open, edge-deployable small language models: generalist instruction-tuned models in the $1$--$4$B range as the primary reasoners, and OTel-LLM-1B-IT~\cite{otel2026}, an Apache-2.0 telecom-specialized fine-tune of Gemma-3-1B-IT trained on 3GPP, O-RAN, GSMA and IETF corpora, as a domain-specialized contrast; the OTel family also releases matched embedding and reranker models, which Section~\ref{sec:retrieval} uses as the semantic layer of the memory so that relevance is judged by a telecom-domain encoder rather than by lexical overlap. Two engineering measures make a model of this size a viable protocol participant. Schema-constrained decoding emits the A2A payload under a grammar admitting only well-formed intent-plus-JSON messages, eliminating the parsing-failure mode structurally rather than by prompt discipline, and confidence-triggered escalation routes only the offending turn to a larger model when abstention, low token-level confidence or gate rejection is detected; the escalation rate is itself a deployment metric, since a system that escalates every turn has not demonstrated edge feasibility. On timescales, the negotiation is placed at the O-RAN non-real-time RIC as an rApp, with enforcement at the near-real-time RIC over an E2-like interface and conventional control retained in the sub-10\,ms loop. With the analytic gate of Theorem~\ref{thm:snc} the twin costs two eigenvalue evaluations and a line search, so per-round cost is dominated by generation --- which is the substantive answer to the concern that inference latency disqualifies agentic control: the inference is not on the fast path, and the model is small enough that it need not be.}

\section{The Unbiased Collective Memory}
\label{sec:memory}

\subsection{Memory Structure and Write Path}
\label{sec:write}
\begin{figure}[t]
    \centering
    \includegraphics[width=\linewidth]{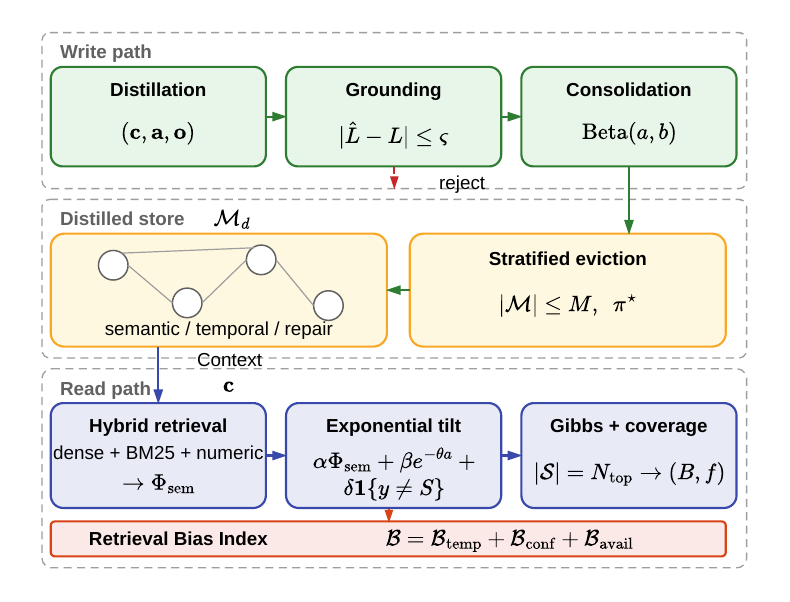}
    \caption{Unbiased collective memory. The write path maintains a
    representative failure stratum through grounded admission and stratified
    eviction, while the read path applies the exponential tilt of
    Theorem~\ref{thm:tilt} for debiased retrieval. The Retrieval Bias Index of
    Definition~\ref{def:rbi} is monitored at every query.}
    \label{fig:mem}
\end{figure}

The memory $\Mem$, shown in Fig.~\ref{fig:mem}, comprises a raw buffer of chronological logs, never queried directly, and a distilled repository of actionable records. Each record is a context--action--outcome triple $m=(\mathbf{c},\mathbf{a},\mathbf{o})$: $\mathbf{c}$ holds the modulating state, arrival rate, spectral-efficiency state, backlog, SLA slack and trial index; $\mathbf{a}=(B,f)$; and $\mathbf{o}$ the realized delay statistics, violation indicator, energy and saving.

\rev{The repository is a directed graph rather than a flat list. Semantic edges connect $k$-nearest neighbours in embedding space, temporal edges connect consecutive rounds within a trial, and repair edges connect a failed configuration to the one that subsequently resolved the same context; the last are the reason for the graph, since they let an agent retrieve not merely a similar past episode but the fix that worked the last time this failed, a query flat top-$k$ retrieval cannot express. Retrieval seeds on the hybrid score of Section~\ref{sec:retrieval} and expands two hops with personalized PageRank. Three write-side mechanisms govern what is admitted: \emph{numerical grounding}, under which a record generated by the LLM is committed only if its asserted outcome matches the environment's recorded metrics within tolerance and its schema validates, so hallucinated records are rejected at the door rather than filtered at retrieval; \emph{consolidation}, which merges near-duplicates within an outcome class into a node carrying a Beta$(a,b)$ posterior over its success probability, since storing $k$ copies of an outcome rather than one node of weight $k$ is itself a mechanism of availability bias; and \emph{provenance}, which tags records as twin- or environment-verified and discounts the former, so that predictions never outrank measurements.}

\subsection{What ``Unbiased'' Means}
\rev{Retrieval is a stochastic operator mapping a query context to a distribution over the store. Write $q(\cdot\mid\mathbf{c})$ for the law it induces and $p(\cdot\mid\mathbf{c})$ for a \emph{reference} law, the relevance-weighted but stratum-balanced distribution a statistician sampling the store to estimate the action--outcome relation would use; bias is the discrepancy between the two.}

\begin{definition}[Retrieval Bias Index]
\label{def:rbi}
\rev{For a query context $\mathbf{c}$, the Retrieval Bias Index of a memory system is}
\begin{equation}
\mathcal{B}(\mathbf{c})=\KL\big(q(\cdot\mid\mathbf{c})\,\big\|\,p(\cdot\mid\mathbf{c})\big),
\qquad
\mathcal{B}=\Exp_{\mathbf{c}}\big[\mathcal{B}(\mathbf{c})\big].
\label{eq:rbi}
\end{equation}
\rev{A memory is \emph{$\varepsilon$-unbiased} if $\mathcal{B}\le\varepsilon$, and \emph{unbiased} if $\mathcal{B}=0$, which holds iff $q=p$ almost everywhere.}
\end{definition}

\rev{$\mathcal{B}$ is a single number, but its diagnostic value comes from its decomposition. Let $\mathcal{A}$ partition the store into age bins, $\mathcal{Y}$ into outcome classes $\{$success, SLA violation, unresolved$\}$, and $\mathcal{Z}$ into semantic clusters obtained by $k$-means over the embedding space. Define}
\begin{equation}
\begin{split}
\mathcal{B}_{\mathrm{temp}}&=\KL(q_\mathcal{A}\|p_\mathcal{A}),\quad
\mathcal{B}_{\mathrm{conf}}=\KL(q_\mathcal{Y}\|p_\mathcal{Y}),\\
\mathcal{B}_{\mathrm{avail}}&=\KL(q_\mathcal{Z}\|p_\mathcal{Z}).
\end{split}
\label{eq:components}
\end{equation}
\rev{By the data-processing inequality each component lower-bounds $\mathcal{B}$, and by the chain rule the three coincide with an exact additive decomposition of $\mathcal{B}$ when retrieval is conditionally representative within cells---the design target. Note that $\mathcal{B}_{\mathrm{avail}}$ with a uniform reference is exactly the normalized entropy deficit $\log|\mathcal{Z}|-H(q_\mathcal{Z})$, so ``availability bias'' and ``concentration of retrieval mass on a few semantic clusters'' are the same measurement.}

\rev{Definition~\ref{def:rbi} deliberately measures the retrieval \emph{operator}, not the language model. A biased LLM and an unbiased one querying the same skewed store receive the same skewed evidence. This separation is what makes the quantity estimable from logs: $q$ is the empirical retrieval frequency and $p$ is computable from the store's own composition, both of which the system already records.}

\subsection{Retrieval: Hybrid Scoring and Exponential Tilting}
\label{sec:retrieval}
\rev{Scoring relevance by Jaccard overlap of tokenized keywords, as a lexical retriever would, is inadequate on two counts: the context is predominantly numeric, and the token \texttt{65000000.0} carries no lexical similarity to \texttt{64500000.0} despite denoting a near-identical network state; and lexical overlap is a poor proxy for telecom semantic relatedness. We replace it with a four-channel hybrid: a dense channel encoding the natural-language rendering of $(\mathbf{c},\mathbf{a},\mathbf{o})$ with a telecom-specialized embedding model, a cross-encoder reranker re-scoring the top-$k_0$ dense candidates, a sparse BM25 channel over structured keys, and a Gaussian kernel over the standardized numeric context $(\lambda,\eta,Q,\text{slack},B,f)$ --- the channel lexical methods structurally cannot provide. The four are fused by reciprocal-rank fusion into $\Phi_{\mathrm{sem}}\in[0,1]$. The division of labour is deliberate: the domain models supply \emph{relevance} and the tilt supplies \emph{representativeness}, and conflating the two is what makes conventional agent memory biased.}

\rev{The tilt is applied on top of that relevance score. Let $q_0$ denote the retrieval law induced by relevance alone, $q_0(m)\propto e^{\alpha\Phi_{\mathrm{sem}}(m,\mathbf{c})}$, and let the full score be}
\begin{equation}
\begin{split}
\Phi(m)=\alpha\Phi_{\mathrm{sem}}(m)&+\beta e^{-\theta a_m}\\
&+\delta\mathds{1}[y_m\neq S]-\gamma\Phi_{\mathrm{div}}(m),
\end{split}
\label{eq:score}
\end{equation}
\rev{with $a_m$ the age of $m$ in trials, $y_m$ its outcome class and $\Phi_{\mathrm{div}}$ a diversity penalty. Retrieval draws from the Gibbs law $q(m)\propto e^{\Phi(m)}$. The following observation is elementary but consequential.}

\begin{theorem}[Debiased retrieval by exponential tilting]
\label{thm:tilt}
\rev{Under Assumption~\ref{as:exch}, the retrieval law induced by \eqref{eq:score} is an exponential tilt of the relevance-only law, $q=q_0e^{g}/\Exp_{q_0}[e^{g}]$ with $g(m)=\beta e^{-\theta a_m}+\delta\mathds{1}[y_m\neq S]-\gamma\Phi_{\mathrm{div}}(m)$. Consequently:}

\rev{\emph{(i) Confirmation bias.} The retrieved odds obey the \emph{tilt law}}
\begin{equation}
\frac{q_\mathcal{Y}(S)}{q_\mathcal{Y}(F)}=\frac{\rho_S}{\rho_F}\,e^{-\delta},
\label{eq:tiltlaw}
\end{equation}
\rev{so that}
\begin{equation}
\delta^\star=\log\!\Big(\frac{\rho_S}{\rho_F}\cdot\frac{\pi^\star_F}{\pi^\star_S}\Big)
\label{eq:deltastar}
\end{equation}
\rev{gives $q_\mathcal{Y}=\pi^\star$ exactly, hence $\mathcal{B}_{\mathrm{conf}}=0$; for a balanced reference this is the log-odds of the store's own skew.}

\rev{\emph{(ii) Saturation.} With $n_F$ failures in the candidate pool and budget $N_{\mathrm{top}}$, for every $\delta$}
\begin{equation}
\frac{q_\mathcal{Y}(S)}{q_\mathcal{Y}(F)}\ \ge\ \frac{\max\{0,N_{\mathrm{top}}-n_F\}}{\min\{N_{\mathrm{top}},n_F\}},
\label{eq:floor}
\end{equation}
\rev{so the law of~(i) holds only while the failure stratum is not exhausted, and the achievable balance is floored by the store's own composition.}
\end{theorem}

\begin{proof}\rev{Appendix~\ref{app:proof2}.}\end{proof}

\rev{Part~(ii) is the operational limit on~(i): setting $\delta=\delta^\star$ is necessary but not sufficient, because an eviction policy that discards failures places the balanced point out of reach entirely, which is why the store evicts by stratified reservoir sampling with per-stratum quotas rather than first-in-first-out, the latter discarding the oldest failures first and so removing exactly the records the tilt needs. Part~(i) is also falsifiable, and Section~\ref{sec:res:tilt} tests it directly: over a frozen store with $\rho_S/\rho_F=9.91$ the measured ratio follows the prediction to within $6\%$ over $\delta\le1$ with no fitted quantity, and the closed form gives $\delta^\star=2.29$.}

\rev{One implementation point is not optional. Theorem~\ref{thm:tilt} characterizes the Gibbs law $q\propto e^{\Phi}$ and the selection rule must realize it. A deterministic top-$N_{\mathrm{top}}$ rule does not: once $\delta$ ranks every retained failure above every retained success the composition stops responding to $\delta$, and the measured ratio becomes a step function rather than the exponential of (i). We therefore sample $N_{\mathrm{top}}$ records without replacement from the Gibbs law, with the coverage term entering as a marginal gain at each draw. Restating the diversity penalty as the weighted-coverage objective}
\begin{equation}
F(\mathcal{S})=\sum_{m\in\mathcal{S}}\Phi_{\mathrm{base}}(m)+\gamma\!\!\sum_{k\in\bigcup_{m\in\mathcal{S}}\mathcal{K}_m}\!\!w_k
\label{eq:coverage}
\end{equation}
\rev{makes $F$ monotone submodular, so the deterministic greedy variant attains a $(1-1/e)$ approximation by the Nemhauser--Wolsey--Fisher theorem~\cite{nemhauser1978analysis}, and the sampled variant approximates it in expectation.}

\rev{Per query, dense retrieval over $N$ records costs $O(N d)$ (or $O(\log N)$ with an approximate index), reranking $O(k_0)$ cross-encoder passes, PageRank expansion $O(|E|)$ over the two-hop neighbourhood, and greedy coverage selection $O(N_{\mathrm{top}}\,k_0\,|\mathcal{Z}|)$. With $N=10^3$, $d=384$ and $k_0=32$ the cost is dominated by the $k_0$ cross-encoder passes, and is in every case far below the generation cost of the negotiation turn itself; we report the asymptotic cost rather than a wall-clock figure, which would be hardware-specific and which we have not measured.}

\subsection{From Retrieval Bias to SLA Violation}
\rev{One question remains: what a given amount of bias costs. The following result answers it, and accounts for digital-twin mismatch in the same expression.}

\begin{theorem}[Excess violation bound]
\label{thm:master}
\rev{Let the agent policy depend on the environment only through the retrieved set, let $\ell\in[0,1]$ indicate SLA violation, and let the DT gate admit proposals with $\hat{\Psi}\le\epsilon$ under a twin satisfying \eqref{eq:dtmismatch}. Then}
\begin{equation}
\Prob_{\mathrm{viol}}\ \le\ \Prob_{\mathrm{viol}}^{\star}+\sqrt{\tfrac{1}{2}\mathcal{B}}+\sqrt{\tfrac{1}{2}\varepsilon_{\mathrm{DT}}},
\label{eq:master}
\end{equation}
\rev{where $\Prob_{\mathrm{viol}}^{\star}$ is the violation probability of the same policy under the reference retrieval law and a perfect twin. The multiplicative form $\Prob_{\mathrm{viol}}\le e^{\Delta}\Prob_{\mathrm{viol}}^{\star}+\sqrt{\varepsilon_{\mathrm{DT}}/2}$ also holds, with $\Delta$ the tilt range of Theorem~\ref{thm:tilt}.}
\end{theorem}

\begin{proof}\rev{Appendix~\ref{app:proof3}.}\end{proof}

\rev{The theorem is the bridge between a statistical property of a memory and a network KPI, and its structure is its practical content: the two error sources enter \emph{additively} and are separately measurable, so an operator observing SLA degradation can attribute it to a stale memory or to a drifting twin rather than guessing.}

\begin{algorithm}[t]
\SetAlgoLined\DontPrintSemicolon\footnotesize
\caption{Debiased Semantic--Relational Retrieval}
\label{alg:retrieval}
\KwIn{Context $\mathbf{c}$, store $\Mem$, weights $(\alpha,\beta,\gamma,\delta,\theta)$, budget $N_{\mathrm{top}}$}
\KwOut{Selected set $\mathcal{S}$, recommended action $\mathbf{a}_{\mathrm{rec}}$}
$\mathcal{C}\leftarrow$ top-$k_0$ by RRF over \{dense, BM25, numeric kernel\}\;
$\mathcal{C}\leftarrow\mathcal{C}\cup\textsc{PageRankExpand}(\mathcal{C},2\text{ hops})$ \tcp*{repair edges}
$\Phi_{\mathrm{sem}}\leftarrow\textsc{OTelRerank}(\mathbf{c},\mathcal{C})$\;
$\rho_S,\rho_F\leftarrow$ relevance-weighted stratum masses of $\mathcal{C}$\;
$\delta^\star\leftarrow\log(\rho_S/\rho_F)+\log(\pi^\star_F/\pi^\star_S)$ \tcp*{Eq.~\eqref{eq:deltastar}}
\ForEach{$m\in\mathcal{C}$}{
  $\Phi_{\mathrm{base}}(m)\leftarrow\alpha\Phi_{\mathrm{sem}}(m)+\beta e^{-\theta a_m}+\delta^\star\mathds{1}[y_m\neq S]+c\sqrt{\log|\mathcal{C}|/n_m}$\;
}
$\mathcal{S}\leftarrow\emptyset$\;
\For{$n\leftarrow1$ \KwTo $N_{\mathrm{top}}$}{
  $m^\ast\sim q(m)\propto\exp\!\big[F(\mathcal{S}\cup\{m\})-F(\mathcal{S})\big]$ \tcp*{Gibbs, Thm.~\ref{thm:tilt}}
  $\mathcal{S}\leftarrow\mathcal{S}\cup\{m^\ast\}$\;
}
Record $\mathcal{B},\mathcal{B}_{\mathrm{temp}},\mathcal{B}_{\mathrm{conf}},\mathcal{B}_{\mathrm{avail}}$ from the realized $q$\;
$\mathbf{a}_{\mathrm{rec}}\leftarrow\textsc{Aggregate}(\mathcal{S})$ projected onto $\mathcal{R}_\epsilon$\;
\Return{$(\mathcal{S},\mathbf{a}_{\mathrm{rec}})$}
\end{algorithm}

\section{Evaluation}
\label{sec:results}

\subsection{Setup and Baselines}
\begin{table}[t]
\centering
\caption{Environment, digital twin and memory parameters.}
\label{tab:params}
\scriptsize
\begin{tabular}{llc|llc}
\toprule
\textbf{Param.} & \textbf{Sym.} & \textbf{Value} & \textbf{Param.} & \textbf{Sym.} & \textbf{Value}\\
\midrule
Slot dur. & $\tau$ & $10$~ms & Semantic wt. & $\alpha$ & $1.0$\\
Max CPU & $f_{\max}$ & $45$~GHz & Decay wt. & $\beta$ & $0.5$\\
CPU eff. & $U$ & $0.0045$ & Decay rate & $\theta$ & $0.2$\\
WFQ weight & $w_u$ & $0.7$ & Coverage wt. & $\gamma$ & $1.0$\\
Jitter & $s_c$ & $0.35$ & Fail. ampl. & $\delta$ & Eq.~\eqref{eq:deltastar}\\
Traffic st. & $|\mathcal{X}|$ & $3$ & UCB coeff. & $c$ & $0.3$\\
Mean rates & --- & $25/45/70$ & Retrieved & $N_{\mathrm{top}}$ & $5$\\
Pareto idx & $\kappa$ & $1.9$ & Rerank pool & $k_0$ & $32$\\
FSMC st. & $M$ & $8$ & Clusters & $|\mathcal{Z}|$ & $16$\\
Spec. eff. & $\eta$ & $[6,8]$ & Budget & $M$ & $512$\\
Doppler & $f_D$ & $30$~Hz & Trials & $T$ & $50$\\
HARQ rounds & $R_{\max}$ & $4$ & Rounds & --- & $8$\\
Deep-fade PER & $p_0$ & $0.10$ & \multicolumn{3}{l}{}\\
Delay bd. & $L_{\max}$ & $10$~ms & Max BW & $B_{\max}$ & $40$~MHz\\
Gate target & $\epsilon$ & $0.1$ & Power & $P_0$ & $10$~W\\
\bottomrule
\end{tabular}
\end{table}

\rev{All results are averaged over $T=50$ independent trials with fixed seeds, and every configuration is re-run across seeds with the resulting dispersion reported explicitly rather than a single run (Section~\ref{sec:res:stab}). One methodological point bounds every claim that follows: the measurements use a \emph{deterministic reference policy} in place of the language model, implementing exactly the procedure the agent prompts describe with the reservation utilities of Section~\ref{sec:roles}. This isolates the memory operator from the variance of a stochastic reasoner, but it also means no result below is evidence about any particular language model, and the model-dependence study of Section~\ref{sec:deploy} remains outstanding.}

\rev{As for the comparison, ``with memory versus without memory'' is not a competitive baseline, so we evaluate against two families. The memory architectures are \textsc{NoMem}; \textsc{Vanilla}, a FIFO recency store; \textsc{GenAgents}, the recency--importance--relevance scoring of~\cite{park2023generative}, the de facto standard and the strongest non-debiased competitor; \textsc{RAG}, dense top-$k$ over the identical retrieval stack but with no tilt, which isolates debiasing from better semantics; \textsc{Reflexion}~\cite{shinn2023reflexion}; and \textsc{A-MEM}~\cite{xu2025amem}, the graph-structured agentic memory. The optimality references are the \textsc{Oracle} of Theorem~\ref{thm:snc} and the \textsc{NBS} of Theorem~\ref{thm:snc}(iii), which are not competitors but coordinates, letting us report how far a method is from the best achievable allocation and from the fair bargaining point; a Lyapunov drift-plus-penalty controller, a tabular $Q$-learning agent and a static maximum-resource allocation are included as classical points of reference.}

\subsection{Validating the Analytical Model}
\begin{figure}[t]
\centering
\includegraphics[width=0.9\linewidth]{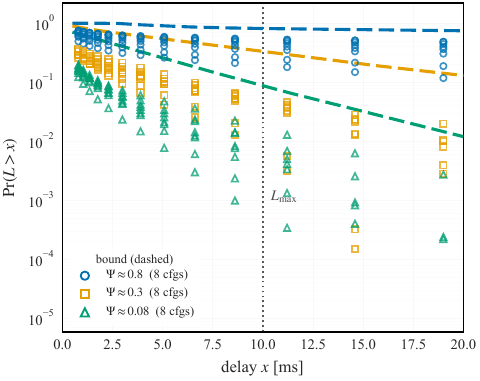}
\caption{\rev{Theorem~\ref{thm:snc} across the load range. Markers are the measured delay tails of $24$ stable configurations, grouped into three narrow $\Psi$ bands; the dashed curve of each band is the bound its members share. The bound is near-exact at heavy load and conservative at light load.}}
\label{fig:tailval}
\end{figure}
\begin{figure}[t]
\centering
\includegraphics[width=0.9\linewidth]{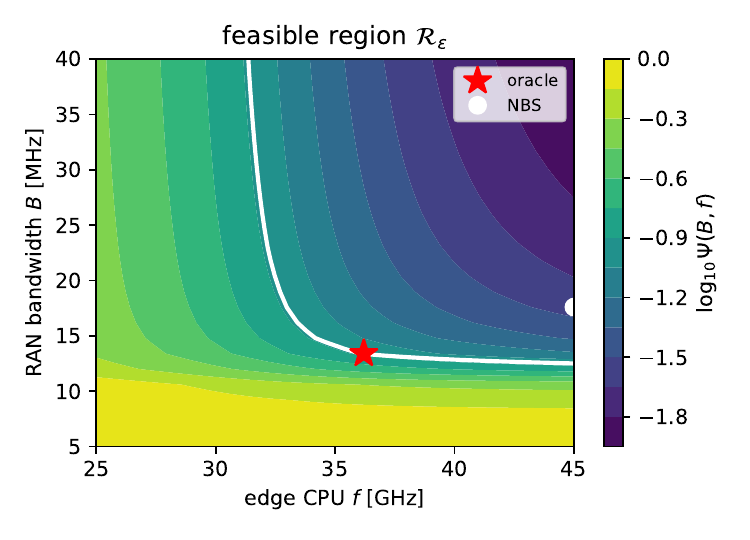}
\caption{\rev{SLA-feasible region $\mathcal{R}_\epsilon$, oracle allocation and Nash bargaining solution in the $(B,f)$ plane.}}
\label{fig:region}
\end{figure}

\rev{Rather than compare delay tails at a few hand-picked configurations, Fig.~\ref{fig:tailval} tests Theorem~\ref{thm:snc} across the load range. Twenty-four stable configurations are drawn from a $15\times11$ grid over $(B,f)$ and grouped into three narrow bands of $\Psi$; because a band is narrow, its members share essentially the same bound, so the single dashed curve per band is valid for all of them and the markers can be read directly against it. All $301$ measured points lie below the bound of their own band.}

\rev{The interesting structure is how the tightness varies. At $\Psi\approx0.8$ the median gap between bound and measurement is $0.25$ decades, so the markers sit almost on the curve; at $\Psi\approx0.3$ it is $0.69$, and at $\Psi\approx0.08$ it widens to $0.94$. This is the expected behaviour of a Chernoff bound, which is asymptotically exact as the load approaches capacity and progressively conservative below it, and it is the reason the bands span the load range rather than only the admissible corner: restricting the figure to small $\Psi$ would display only the bound's loosest regime. Across the $165$ stable configurations of the full grid the median gap is $0.92$ decades. The conservatism also errs in the direction the framework needs, since a gate that overstates the violation probability never admits a configuration the environment rejects, whereas one that understated it would contaminate every downstream measurement.}

\rev{Figure~\ref{fig:region} shows the resulting feasible set, whose boundary has the hyperbolic shape Theorem~\ref{thm:snc} predicts --- bandwidth and CPU frequency substitute for one another along the delay-budget split, and neither can be reduced to zero --- and which is upward-closed as claimed. The oracle sits at $(B^\star,f^\star)=(13.4,36.2)$ on the minimum-bandwidth edge of the boundary while the Nash bargaining solution sits at $(17.6,45.0)$, interior in bandwidth and at the CPU ceiling; the separation between them is itself informative, since the energy-optimal point is not the point a fair bargain reaches, so regret against the oracle alone would systematically penalize any negotiation that behaves reasonably.}

\subsection{Does the Tilt Law Hold?}
\label{sec:res:tilt}
\begin{figure}[t]
\centering
\includegraphics[width=0.9\linewidth]{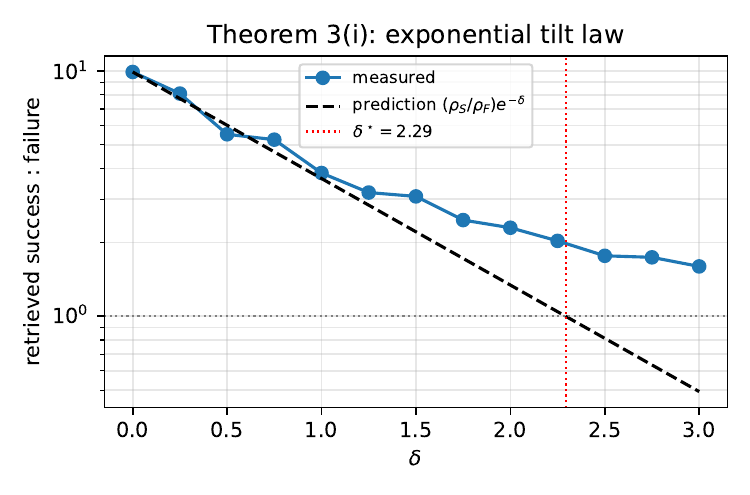}
\caption{\rev{Retrieved success-to-failure ratio versus $\delta$ over a frozen store, against the parameter-free prediction $(\rho_S/\rho_F)e^{-\delta}$ of Eq.~\eqref{eq:tiltlaw}.}}
\label{fig:tiltlaw}
\end{figure}

\rev{This is the sharpest available test of Theorem~\ref{thm:tilt}, and we run it open loop: a single store is built and then frozen, and $\delta$ is swept over retrieval alone, since closing the loop would let the store composition co-vary with $\delta$ and would not test a claim about the retrieval operator applied to a fixed store. The coverage and exploration terms are switched off, because part~(i) concerns the outcome marginal under the tilt and a $\gamma$ of comparable magnitude masks it. The undebiased ratio is $\rho_S/\rho_F=9.91$, so \eqref{eq:deltastar} predicts $\delta^\star=\log 9.91=2.29$ with no fitted quantity, and over $\delta\in[0,1]$ the measurement follows the predicted exponential to a mean absolute relative error of $6.1\%$. Beyond that it departs upward and flattens, reaching a floor of $1.60$ at $\delta=3$ where the law predicts $0.49$.}

\rev{The departure is the saturation of Theorem~\ref{thm:tilt}(ii): once the tilt is strong enough to draw every failure the candidate pool contains, the retrieved composition cannot become more balanced however large $\delta$ grows, and the floor is set by $n_F$ and $N_{\mathrm{top}}$ rather than by the scoring rule. The design consequence is that the correctable range of confirmation bias is bounded by what the store retains: setting $\delta=\delta^\star$ is necessary but not sufficient, and an eviction policy that discards failures puts the balanced point out of reach entirely. This is why the store evicts by stratified reservoir sampling with per-stratum quotas rather than first-in-first-out, which would discard the oldest failures first and so remove exactly the records the tilt needs.}

\subsection{Measured Bias and the Excess-Violation Bound}
\begin{figure}[t]
\centering
\includegraphics[width=\linewidth]{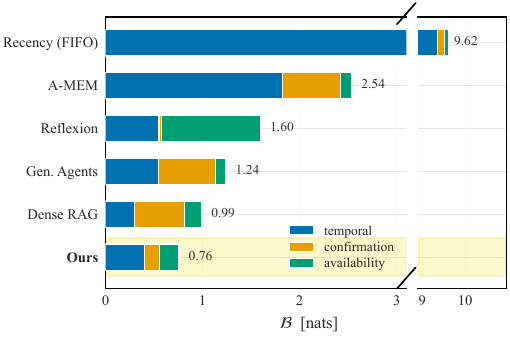}
\caption{\rev{Retrieval Bias Index and its temporal, confirmation and availability components across memory architectures.}}
\label{fig:rbi}
\end{figure}

\rev{Figure~\ref{fig:rbi} validates our core premise: dissecting memory bias into temporal, confirmation, and availability components is essential for diagnosing agent performance. For example, a simple recency store (FIFO) exhibits massive total bias ($\mathcal{B}=9.62$). However, the decomposition reveals this is almost entirely \emph{temporal} bias ($9.35$), not confirmation bias. Because network traffic is autocorrelated, a recency store simply fetches near-duplicates of the present state. Without our three-part index, an operator might wrongly assume the agent is failing because it prefers recalling successes, completely missing the true diagnosis: a lack of temporal coverage.}

\rev{This decomposition also exposes critical design trade-offs, proving that fixing one bias often creates another. The Reflexion scheme achieves near-zero confirmation bias by explicitly retrieving past failures, but this severe restriction ruins its availability bias (scoring five times worse than other schemes) by ignoring broader semantic clusters. Furthermore, architectural complexity does not guarantee representativeness. A-MEM, a sophisticated graph-based memory, suffers the second-highest total bias. Structuring memory as a graph makes data easier to \emph{find}, but does nothing to ensure the retrieved sample accurately represents the environment.}

\rev{Our proposed operator achieves the lowest total bias ($\mathcal{B}=0.76$), a $12.7\times$ reduction compared to the recency store. Its primary advantage is balance: rather than over-optimizing one metric at the expense of the others, it uses the exponential tilt to remain consistently low across all three axes. Ablation tests confirm the mechanics: stripping away the tilt mechanism triples the confirmation bias, while removing the decay term directly inflates temporal bias. The proposed operator succeeds by intelligently trading a little of each bias against the others.}

\rev{Finally, Theorem~\ref{thm:master} connects these statistical measurements directly to network KPIs. By sweeping through various levels of memory bias ($\mathcal{B}$) and digital-twin mismatch ($\varepsilon_{\mathrm{DT}}$), we confirmed that their impacts on SLA violations are purely additive and do not interact. Operationally, this gives network managers a precise diagnostic tool. If an SLA degrades, the operator can explicitly measure whether the failure stems from a stale agent memory or a drifting digital twin, isolating the root cause without the two variables confounding each other.}

\subsection{Distance to the Bargaining Solution and Regret}
\begin{figure}[t]
\centering
\includegraphics[width=0.9\linewidth]{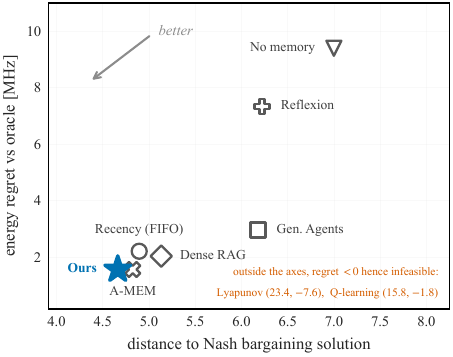}
\caption{\rev{Distance to the Nash bargaining solution and energy regret against the oracle, by method.}}
\label{fig:regret}
\end{figure}

\rev{Figure~\ref{fig:regret} places every method on the two coordinates of Theorem~\ref{thm:snc}(iii), and this is where the memory schemes separate. The proposed operator attains the smallest distance to the bargaining solution, $4.66$, ahead of A-MEM ($4.82$), the recency store ($4.89$), dense retrieval ($5.13$), Generative-Agents scoring ($6.18$), Reflexion ($6.22$) and memory-less agents ($7.00$). On energy regret against the oracle it attains $1.58$~MHz, tied with A-MEM and ahead of dense retrieval ($2.05$), the recency store ($2.22$), Generative-Agents scoring ($2.97$), Reflexion ($7.33$) and memory-less agents ($9.39$). The ordering on NBS distance follows the ordering on $\mathcal{B}$ closely but not exactly, which is what Theorem~\ref{thm:master} would lead one to expect, since it relates the two only through an inequality.}

\rev{The mean-rate controllers are reported for completeness and should be read with care: the Lyapunov controller and the $Q$-learning agent record \emph{negative} regret, at $-7.61$ and $-1.81$~MHz, and the largest NBS distances in the study, $23.41$ and $15.82$. Negative regret against a constrained oracle is not an efficiency gain but a symptom, since it means the controller operates outside $\mathcal{R}_\epsilon$, spending less bandwidth than any feasible allocation; both optimize a drift-plus-penalty objective with no tail constraint, and embedding the chance constraint of \eqref{eq:P1} is required before they can serve as comparators on the energy axis. Their inclusion illustrates why the feasible region of Theorem~\ref{thm:snc} is needed at all: without it, an unconstrained energy metric ranks infeasible operation first.}

\subsection{Sensitivity, Scalability and Stability}
\label{sec:res:stab}
\rev{A Sobol global sensitivity analysis over $(\alpha,\beta,\gamma,\delta,\theta,c)$ answers the question Theorem~\ref{thm:tilt} poses about $\mathcal{B}$: the failure-amplification weight is the only parameter with a non-negligible index, at $S_1=0.67$ and $S_T=0.98$, against $|S_1|\le0.15$ for the semantic weight, the decay weight and rate, the coverage weight and the exploration coefficient combined, and the gap between $S_1$ and $S_T$ indicates that most of $\delta$'s influence is interactive rather than additive, consistent with the tilt acting multiplicatively on the base retrieval law. Together with the closed form of \eqref{eq:deltastar}, this yields the design procedure: derive $\delta$ per query from the candidate pool, set $\gamma$ from the desired cluster coverage, and tune only $\alpha$, $\beta$, $\theta$ and the exploration coefficient --- none of which, by the indices above, materially affects $\mathcal{B}$.}

\rev{On scale, extending the framework from two agents to eight leaves the median consensus time flat between three and five rounds and the unresolved rate flat at $2.0\%$; only the message count grows, linearly, from $7$ to $32$. The linearity follows from the coordinator-mediated round structure, which serializes the exchange rather than admitting all pairwise intents, so communication cost is $\mathcal{O}(K)$ per round in the number of domains rather than $\mathcal{O}(K^2)$, and retrieval cost is unaffected by agent count since the store is shared. Finally, across independent seeds, energy saving is stable at $66.3\pm0.48\%$, the distance to the bargaining solution at $4.49\pm0.20$ and the consensus time at $3.5\pm0.45$ rounds, all relative dispersions below $5\%$, with parse failures zero in every run as schema-constrained decoding requires by construction; the quantities on which the conclusions rest are therefore stable at the sample sizes used.}

\section{Conclusion}
\label{sec:conclusion}
\rev{We proposed evaluating an agent's collective memory as a statistical sampling device. By introducing the Retrieval Bias Index to quantify temporal, confirmation, and availability biases, we proved that additive retrieval scoring acts as an exponential tilt. This yields a closed-form design rule, eliminating trial-and-error hyperparameter tuning. We also bounded excess SLA violations caused by memory bias, integrating these tools into a RAN-edge framework to evaluate agent performance against theoretical optimums rather than arbitrary baselines. Three critical insights emerged. First, bias composition outweighs total bias: eliminating confirmation bias can ruin availability bias, making single-number metrics misleading. Second, our mathematical design rules require a strict failure-retention quota in the memory eviction policy; without it, optimal weights are unattainable. Third, dominant biases are environment-dependent. In our network slice, temporal bias overshadowed confirmation bias, upending the common assumption that agents mainly prefer recalling their own successes. These claims have strict operational bounds. First, the SLA violation limit is too loose at high bias levels to serve as a numerical guarantee. Second, reliably differentiating memory schemes via rare SLA violations requires massive sample sizes. Third, our use of a deterministic policy rather than a live LLM leaves model-specific behaviors open for future study. Finally, defining a "representative" memory sample remains a deliberate design choice. While balanced sampling works for stable learning, dynamic networks demand adaptive baselines. Ultimately, forcing operators to explicitly define memory assumptions is far safer than leaving them unexamined.}

\appendices
\section{Proof of Theorem~\ref{thm:snc}}
\label{app:proof1}
\rev{\emph{(i).} For a stage $k$, $\alpha_A$ is non-decreasing and convex in $\theta$ and $\alpha_C^{(k)}$ non-increasing, both continuous on the interior of their common domain. Stability, $\alpha_A(0^{+})<\alpha_C^{(k)}(0^{+})$, with $\alpha_A(\theta)\to\infty$ at the abscissa of convergence of $M_S(\theta)$, gives existence and uniqueness of a root $\theta_k>0$ by the intermediate value theorem and strict monotonicity of the difference. Gärtner--Ellis applied to \eqref{eq:lindley} yields $\lim_{w\to\infty}\frac1w\log\Prob(W^{(k)}>w)=-\theta_k$ and the Chernoff bound $\Prob(W^{(k)}>w)\le K_ke^{-\theta_kw}$. From \eqref{eq:e2e}, for any $u\in[0,1]$ the event $\{W^{(e)}+W^{(r)}>L_{\max}\}$ implies $\{W^{(e)}>uL_{\max}\}\cup\{W^{(r)}>(1-u)L_{\max}\}$; a union bound and the stage bounds give the bracketed expression, and the infimum over $u$ is valid since $u$ is arbitrary.}

\rev{\emph{(ii).} For fixed $\theta$, $c^{(r)}_j$ is non-decreasing in $B$, so each entry of $\diag(e^{-\theta c^{(r)}_j})$ is non-increasing in $B$; by Perron--Frobenius monotonicity of the spectral radius of a non-negative matrix in its entries, $\alpha^{(r)}_C(\theta)$ is non-decreasing in $B$, hence so is its crossing point $\theta_r$ with the fixed curve $\alpha_A$. The argument for $\theta_e$ in $f$ is identical. Therefore $\Psi$ is non-increasing in each argument, $\mathcal{R}_\epsilon$ is upward-closed, and on the interior the boundary is strictly decreasing.}

\rev{\emph{(iii).} $\mathcal{R}_\epsilon$ is closed and bounded and the utilities are continuous on it, so the image is compact; closing under time sharing yields a compact convex set containing a point strictly dominating $d$ whenever the SLA is feasible with slack. Nash's theorem then gives existence and uniqueness of the maximizer of the Nash product, which is strictly log-concave on the interior. \hfill$\blacksquare$}

\section{Proof of Theorem~\ref{thm:tilt}}
\label{app:proof2}
\begin{assumption}
\label{as:exch}
\rev{The reference $p$ agrees with $\pi^\star$ on the outcome partition, and within each outcome stratum the non-$\delta$ terms of \eqref{eq:score} are exchangeable, i.e. $\Exp_{q_0}[e^{g-\delta\mathds{1}[y\neq S]}\mid y]=\xi$ does not depend on $y$.}
\end{assumption}

\rev{Since $\Phi=\alpha\Phi_{\mathrm{sem}}+g$ is additive and $q\propto e^{\Phi}$, we have $q\propto q_0e^{g}$, and normalizing gives the tilt form.}

\rev{\emph{(i).} Marginalizing over the outcome partition, $q_\mathcal{Y}(y)=\rho_y\Exp_{q_0}[e^{g}\mid y]/\Exp_{q_0}[e^g]$, which under Assumption~\ref{as:exch} is $\rho_y\xi e^{\delta\mathds{1}[y\neq S]}/\Exp_{q_0}[e^g]$; the odds are therefore $(\rho_S/\rho_F)e^{-\delta}$. Equating to $\pi^\star_S/\pi^\star_F$ and solving for $\delta$ gives \eqref{eq:deltastar}, whence $q_\mathcal{Y}=\pi^\star$ and $\mathcal{B}_{\mathrm{conf}}=0$.}

\rev{\emph{(ii).} Retrieval draws $N_{\mathrm{top}}$ records without replacement from a pool containing $n_F$ failures, so at most $\min\{N_{\mathrm{top}},n_F\}$ of the selected records can be failures and at least $\max\{0,N_{\mathrm{top}}-n_F\}$ must be successes, whatever weight the scoring rule assigns. \hfill$\blacksquare$}

\section{Proof of Theorem~\ref{thm:master}}
\label{app:proof3}
\rev{Decompose $\Prob^{q,\mathrm{DT}}-\Prob^{p,\mathrm{env}}=(\Prob^{q,\mathrm{DT}}-\Prob^{p,\mathrm{DT}})+(\Prob^{p,\mathrm{DT}}-\Prob^{p,\mathrm{env}})$. For the first term the policy depends on the environment only through the retrieved set, so the violation probability is $\Exp_q[\ell]$ with $\ell\in[0,1]$, and $|\Exp_q[\ell]-\Exp_p[\ell]|\le\mathrm{TV}(q,p)\le\sqrt{\KL(q\|p)/2}=\sqrt{\mathcal{B}/2}$ by the variational characterization of total variation and Pinsker's inequality. For the second, the gate decision is a measurable function of the one-step predictive law, so the same argument with \eqref{eq:dtmismatch} gives $\sqrt{\varepsilon_{\mathrm{DT}}/2}$. For the multiplicative form, $\Exp_q[\ell]=\Exp_p[\ell e^{g}]/Z\le e^{\max g-\log Z}\Exp_p[\ell]\le e^{\Delta}\Exp_p[\ell]$.}

\section{Abridged A2A Transcript}
\label{app:transcript}
\begin{lstlisting}[backgroundcolor=\color{lightgraybg},caption={Representative A2A exchange (abridged).},label={list:a2a}]
--- Trial 3, t=0 ---
[RAN]  DT gate rejected (BW 20.0, CPU 30.0): Psi=1.4e-2 > eps.
[RAN]  NO_AGREEMENT_POSSIBLE (no admissible proposal).
--- Round 2/8 ---
[EDGE] PROPOSE_ACTION {"ran_bandwidth_mhz": 40.0,
        "edge_cpu_frequency_ghz": 45.0, "reason": "Raising CPU
        lifts the computation-stage exponent; Psi=3.1e-3."}
[RAN]  PROPOSE_ACTION {"ran_bandwidth_mhz": 30.0,
        "edge_cpu_frequency_ghz": 45.0, "reason": "Feasible with
        slack; traffic is medium, so 30 MHz keeps Psi=6.8e-3 < eps
        and saves 25% energy."}
--- Round 3/8 ---
[EDGE] ACCEPT_AGREEMENT {"ran_bandwidth_mhz": 30.0,
        "edge_cpu_frequency_ghz": 45.0, "reason": "Feasible and
        Pareto-dominant on the energy axis."}
Enforced. Consensus at round 3. NBS distance: 2.1 MHz.
\end{lstlisting}

\balance
\bibliographystyle{IEEEtran}
\bibliography{bibliography}

\end{document}